\documentclass[aps,prc,superscriptaddress,amsfonts,amssymb,amsmath,bibnotes,footinbib,twocolumn,fleqn,10pt]{revtex4-1}

\usepackage{graphicx}
\usepackage{subfigure}
\usepackage{leftidx}
\usepackage{setspace}
\usepackage{array}
\usepackage{lineno,hyperref}
\usepackage{amssymb}
\usepackage{amsmath}
\usepackage[OT2,T1]{fontenc}
\DeclareSymbolFont{cyrletters}{OT2}{wncyr}{m}{n}
\DeclareMathSymbol{\Pma}{\mathalpha}{cyrletters}{"50}
\DeclareMathSymbol{\ima}{\mathalpha}{cyrletters}{"69}
\DeclareMathSymbol{\sma}{\mathalpha}{cyrletters}{"73}
\DeclareMathSymbol{\Mma}{\mathalpha}{cyrletters}{"6D}
\DeclareMathSymbol{\ama}{\mathalpha}{cyrletters}{"61}
\DeclareMathSymbol{\Bma}{\mathalpha}{cyrletters}{"76}
\DeclareMathSymbol{\Ama}{\mathalpha}{cyrletters}{"41}
\DeclareMathSymbol{\Yma}{\mathalpha}{cyrletters}{"59}
\DeclareMathSymbol{\bma}{\mathalpha}{cyrletters}{"5C}
\DeclareMathSymbol{\bmb}{\mathalpha}{cyrletters}{"7E}
\DeclareMathSymbol{\bmc}{\mathalpha}{cyrletters}{"7B}
\DeclareMathSymbol{\bmd}{\mathalpha}{cyrletters}{"7D}
\DeclareMathSymbol{\Qma}{\mathalpha}{cyrletters}{"51}
\DeclareMathSymbol{\Cma}{\mathalpha}{cyrletters}{"5C}
\DeclareMathSymbol{\Cmb}{\mathalpha}{cyrletters}{"63}
\DeclareMathSymbol{\Cmc}{\mathalpha}{cyrletters}{"68}
\DeclareMathSymbol{\Cmd}{\mathalpha}{cyrletters}{"61}
\DeclareMathSymbol{\Cme}{\mathalpha}{cyrletters}{"72}
\DeclareMathSymbol{\Cmf}{\mathalpha}{cyrletters}{"27}
\DeclareMathSymbol{\Cmg}{\mathalpha}{cyrletters}{"30}
\DeclareMathSymbol{\Cmh}{\mathalpha}{cyrletters}{"30}
\DeclareMathSymbol{\Cmi}{\mathalpha}{cyrletters}{"33}
\DeclareMathSymbol{\Zma}{\mathalpha}{cyrletters}{"5A}

\begin{document}
\begin{spacing}{1.0}

\title{Observation of $^{6}$He+$t$ cluster states in  $^{9}$Li}



\author{W.~H.~Ma}
\affiliation{Key Laboratory of High Precision Nuclear Spectroscopy and Center for Nuclear Matter Science, Institute of Modern Physics, Chinese Academy of Science, Lanzhou 730000, People's Reublic of China}
\affiliation{University of Chinese Academy of Science, Beijing, 100049, People's Reublic of China}
\affiliation{Lanzhou University, Lanzhou 730000, China}
\affiliation{Fudan University, Shanghai 200433, China}
\author{J.~S.~Wang}
\email[]{jswang@impcas.ac.cn}
\affiliation{Key Laboratory of High Precision Nuclear Spectroscopy and Center for Nuclear Matter Science, Institute of Modern Physics, Chinese Academy of Science, Lanzhou 730000, People's Reublic of China}
\affiliation{University of Chinese Academy of Science, Beijing, 100049, People's Reublic of China}
\author{D.~Patel}
\affiliation{Key Laboratory of High Precision Nuclear Spectroscopy and Center for Nuclear Matter Science, Institute of Modern Physics, Chinese Academy of Science, Lanzhou 730000, People's Reublic of China}
\author{Y.~Y.~Yang}
\affiliation{Key Laboratory of High Precision Nuclear Spectroscopy and Center for Nuclear Matter Science, Institute of Modern Physics, Chinese Academy of Science, Lanzhou 730000, People's Reublic of China}
\author{J.~B.~Ma}
\affiliation{Key Laboratory of High Precision Nuclear Spectroscopy and Center for Nuclear Matter Science, Institute of Modern Physics, Chinese Academy of Science, Lanzhou 730000, People's Reublic of China}
\author{S.~L.~Jin}
\affiliation{Key Laboratory of High Precision Nuclear Spectroscopy and Center for Nuclear Matter Science, Institute of Modern Physics, Chinese Academy of Science, Lanzhou 730000, People's Reublic of China}
\author{P.~Ma}
\affiliation{Key Laboratory of High Precision Nuclear Spectroscopy and Center for Nuclear Matter Science, Institute of Modern Physics, Chinese Academy of Science, Lanzhou 730000, People's Reublic of China}
\author{Q.~Hu}
\affiliation{Key Laboratory of High Precision Nuclear Spectroscopy and Center for Nuclear Matter Science, Institute of Modern Physics, Chinese Academy of Science, Lanzhou 730000, People's Reublic of China}
\author{Z.~Bai}
\affiliation{Key Laboratory of High Precision Nuclear Spectroscopy and Center for Nuclear Matter Science, Institute of Modern Physics, Chinese Academy of Science, Lanzhou 730000, People's Reublic of China}
\author{M.~R.~Huang}
\affiliation{Key Laboratory of High Precision Nuclear Spectroscopy and Center for Nuclear Matter Science, Institute of Modern Physics, Chinese Academy of Science, Lanzhou 730000, People's Reublic of China}
\author{X.~Q.~Liu}
\affiliation{Key Laboratory of High Precision Nuclear Spectroscopy and Center for Nuclear Matter Science, Institute of Modern Physics, Chinese Academy of Science, Lanzhou 730000, People's Reublic of China}
\author{Y.~J.~Zhou}
\affiliation{Key Laboratory of High Precision Nuclear Spectroscopy and Center for Nuclear Matter Science, Institute of Modern Physics, Chinese Academy of Science, Lanzhou 730000, People's Reublic of China}
\affiliation{University of Chinese Academy of Science, Beijing, 100049, People's Reublic of China}
\author{J.~Chen}
\affiliation{Key Laboratory of High Precision Nuclear Spectroscopy and Center for Nuclear Matter Science, Institute of Modern Physics, Chinese Academy of Science, Lanzhou 730000, People's Reublic of China}
\affiliation{University of Chinese Academy of Science, Beijing, 100049, People's Reublic of China}
\author{Z.~H.~Gao}
\affiliation{Key Laboratory of High Precision Nuclear Spectroscopy and Center for Nuclear Matter Science, Institute of Modern Physics, Chinese Academy of Science, Lanzhou 730000, People's Reublic of China}
\author{Q.~Wang}
\affiliation{Key Laboratory of High Precision Nuclear Spectroscopy and Center for Nuclear Matter Science, Institute of Modern Physics, Chinese Academy of Science, Lanzhou 730000, People's Reublic of China}
\author{J.~Lubian}
\affiliation{ Instituto de F$\imath$sica, Universidade Federal Fluminense, Av. Litoranea s/n, Gragoat$\acute{a}$, Niteroi, R.J., 24210-340, Brazil.}
\author{J.~X.~Li}
\affiliation{Southwest University, Chongqing 400044, People's Reublic of China}
\author{T.~F.~Wang}
\affiliation{BeiHang University, Beijing 100083, People's Reublic of China}
\author{S.~Mukherjee}
\affiliation{Physics Department, Faculty of Science, M.S. University of Baroda, Vadodara - 390002, India}
\author{X.~Y.~Ju}
\affiliation{University of Science and Technology of China, Hefei 230026, People's Reublic of China}
\author{Y.~S.~Yu}
\affiliation{Chongqing University, Chongqing 400044, People's Reublic of China}
\author{T.~W.~Wu}
\affiliation{BeiHang University, Beijing 100083, People's Reublic of China}
\author{C.~Ni}
\affiliation{BeiHang University, Beijing 100083, People's Reublic of China}
\author{X.~D.~Jia}
\affiliation{Chongqing University, Chongqing 400044, People's Reublic of China}
\author{Q.~B.~Liu}
\affiliation{Chongqing University, Chongqing 400044, People's Reublic of China}
\author{Y.~H.~Zhang}
\affiliation{Key Laboratory of High Precision Nuclear Spectroscopy and Center for Nuclear Matter Science, Institute of Modern Physics, Chinese Academy of Science, Lanzhou 730000, People's Reublic of China}
\author{H.~S.~Xu}
\affiliation{Key Laboratory of High Precision Nuclear Spectroscopy and Center for Nuclear Matter Science, Institute of Modern Physics, Chinese Academy of Science, Lanzhou 730000, People's Reublic of China}
\author{G.~Q.~Xiao}
\affiliation{Key Laboratory of High Precision Nuclear Spectroscopy and Center for Nuclear Matter Science, Institute of Modern Physics, Chinese Academy of Science, Lanzhou 730000, People's Reublic of China}



\date{\today}

\begin{abstract}
$^{6}$He+$t$ cluster states of exited $^{9}$Li have been measured by 32.7 MeV/nucleon $^{9}$Li beams bombarding on $^{208}$Pb target. Two resonant states are clearly observed with the excitation energies at 9.8 MeV and 12.6 MeV and spin-parity of 3/2$^{-}$ and  7/2$^{-}$ respectively. These two states are considered to be members of K$^{\pi}$=1/2$^{-}$ band. The spin-parity of them are identified by the method of angular correlation analysis and verified by the continuum discretized coupled channels (CDCC) calculation, which agrees with the prediction of the generator coordinate method (GCM). A monopole matrix element about 4 fm$^{2}$ for the  3/2$^{-}$ state is extracted from the distorted wave Born approximation (DWBA) calculation. These results strongly support the feature of clustering structure of two neutron-rich clusters in the neutron-rich nucleus $^{9}$Li for the first time.

\end{abstract}


\maketitle

\section{INTRODUCTION}
Clustering is an interesting phenomenon existing in many scientific fields from the macroscale world, such as galaxy in the universe, to the microscale world, such as pentaquarks in the quark matter. Nucleon clustering in a nucleus has been proposed since 1937 \cite{bibitem1}. Alpha particle is easy to form a cluster in a nucleus because of its high stability and a strong and repulsive alpha-alpha interaction \cite{bibitem2}. One of the famous cluster structure is Hoyle state due to its significance on the massive $^{12}$C production in the Universe \cite{bibitem3}.

With the development of Radioactive Isotope Beam facilities, clustering of light nuclei has been discovered as one of the unique quantal features. Many theoretical and experimental efforts are focused on the clustering of neutron-excess Be and C isotopes. A well-established $\alpha$+$\alpha$ rotor of $^{8}$Be is a representative example of clustering structure \cite{bibitem4}. In highly excited states near the He+He threshold energy of $^{10}$Be \cite{bibitem5,bibitem6,bibitem7,bibitem8} and $^{12}$Be \cite{bibitem9,bibitem10,bibitem11}, well-developed cluster states have been extensively studied both in experimental and theoretical sides in recent decades. The $\alpha$ cluster plays a critical role in clustering of Be and C isotopes.

Based on the Ikeda's threshold rule \cite{bibitem12}, the excitation energy of a cluster state is rather close to the cluster-decay threshold energy. The clustering states can be studied from the resonance decay spectroscopy by detecting all of the decay fragments and the excitation energy of the parent state can be determined from the invariant mass of the detected decay products \cite{bibitem13}.

Many investigations of cluster formation for the Li isotopes have been done in last decades, where the unstable $t$ cluster, or together with $\alpha$ cluster, plays a critical role. Clustering structures, such as $\alpha$+$d$ and $^{3}$He+$t$ in $^{6}$Li and $\alpha$+$t$ in $^{7}$Li have been studied both in theory and experiment \cite{bibitem14,bibitem15,bibitem16,bibitem17,bibitem18,bibitem19}.

For $^{9}$Li, many theoretical descriptions of the cluster structures have been studied but experimental investigations are scarce. The no-core shell model calculation and the tensor optimized shell model calculation are feasible for the low-lying states \cite{bibitem20,bibitem21}. Additionally, a $\alpha$+$t$+$n$+$n$ cluster model is used to describe the lowlying states of $^{9}$Li nucleus\cite{bibitem22}. The stochastic multiconfiguration mixing method with the $\alpha$+$t$+$n$+$n$ and $t$+$t$+$t$ configurations has predicted the exotic excited state with three triton clusters \cite{bibitem23}. With the quadrupole deformation ($\beta$-$\gamma$) constraint in the framework of antisymmetrized molecular dynamics (AMD), a largely deformed state having a $^{6}$He+t structure appear in excited states of $^{9}$Li is shown \cite{bibitem24}.  In Ref. \cite{bibitem25}, The $^{6}$He+$t$ cluster structure has also been predicted with generator coordinate method (GCM) calculation. The results suggest that $^{6}$He+$t$ cluster states near the $^{6}$He+$t$ threshold energy may construct a K$^{\pi}$ = $1/2^{-}$ band.

Furthermore, Isoscalar monopole transitions from the ground states to cluster states in $^{10}$Be and $^{9}$Li are investigated using $^{6}$He+$\alpha$ and $^{6}$He+$t$ cluster models, respectively \cite{bibitem26}.  As pointed out in Refs. \cite{bibitem27,bibitem28,bibitem29,bibitem6,bibitem30}, a strong monopole strength comparable to the typical single-particle strength for excited states below 20 MeV has been proposed as a sensitive probe for cluster formation and can be observed experimentally. Moreover, the studies in Refs. \cite{bibitem13,bibitem31}, show that the monopole transition strength is also a promising tool to signal the cluster formation in unstable nuclei.

However, the properties of $^{9}$Li are not known well and very few number of experiments have been carried out to investigate the cluster structures and clustering states. It is very important for understanding the cluster formation to search for such resonance states of two neutron-rich clusters ($^{6}$He and $t$) in the unstable nucleus $^{9}$Li. In the present work, we investigate the $^{6}$He+$t$ cluster states for the first time by implementing coincidence measurements between the ontgoing $^{6}$He+$t$ fragments at the most forward angles in the reaction of $^{9}$Li with $^{208}$Pb target. The angular correlation analysis, GCM calculation and continuum discretized coupled channels (CDCC) calculation are used to identify these cluster states. A monopole strength comparable to the typical single-particle strength and a wide decay width are extracted for the 3/2$^{-}$ resonant state, which demonstrates a typical clustering structure in $^{9}$Li. The Sec. II of this paper is devoted to the description of the experimental facility and measurements. The Sec. III is dedicated to the analysis of the experimental data by mean of the CDCC calculations. Finally, in Sec. IV we give the main conclusions of the present work.

\section{EXPERIMENTAL DETAILS}
The experiment was performed at the Heavy Ion Research Facility in Lanzhou (HIRFL). The primary beams, $^{12}$C with the energy of 53.7 MeV/nucleon, were delivered by the HIRFL and bombarded on the production target of 3038 mm $^{9}$Be to produce the secondary beams. The 32.7 MeV/nucleon secondary $^{9}$Li beams were separated and purified with a 2153 mm  Al degrader by Radioactive Ion Beam Line in Lanzhou (RIBLL) \cite{bibitem32}. The intensity of the secondary $^{9}$Li beam was about $1.1\times10^{3}$ particles per second with 99$\%$ purity. A self-supported natural Pb target with a thickness of 526.9 mg/cm$^{2}$ was used as a reaction target. Three parallel-plate avalanche counters (PPACs) \cite{bibitem33} with position resolutions better than 1 mm were placed before the reaction target to determine the position and incident angle of the secondary beam event by event.  A zero-degree  telescope array, consisted of two $\Delta$E  detectors (double-sided silicon strip detectors, DSSDs) and a E detector (8$\times$8 CsI(Tl) scintillator array) covering the $\theta$ angles from 0$^{\circ}$ to 10$^{\circ}$, was used to measure the charged fragments. The two DSSDs, with a thickness of 523 $\mu$m for the DSSD1 and 527 $\mu$m for the DSSD2 respectively, have the same sensitive area of 49$\times$49 mm$^{2}$. Each DSSD is divided into 16 strips both in the front and rear sides; the width of each strip is 3 mm and the interval is 0.1 mm. Each CsI(Tl) scintillator is composed by a trustum of a pyramid with 21$\times$21 mm$^{2}$ in front side, 23$\times$23 mm$^{2}$ in back side and the length of each scintillator is 50 mm. Each scintillator is coupled with a Photomultiplier tube (PMT). A description of this kind of telescope array can also be found in Refs. \cite{bibitem34,bibitem35}. The secondary beams of $^{9}$Li, $^{6}$He, $^{4}$He, and $^{3}$H produced from the $^{12}$C primary beam were used to calibrate the telescope. More details for the experimental setup and measurement can be found in our previous publication \cite{bibitem36}.

All charged particles are identified by the telescope array. The events with $t$ and $^{6}$He, which are decayed from the excited $^{9}$Li and recorded in the telescope, are selected. The total kinetic energies of these two particles are obtained by summing the energy loss in the two DSSDs and the residue energy in CsI(Tl) scintillator. Their tracks are recorded by the two DSSDs. Thus, the relative energy $E_{rel}$ of a pair of fragments can be deduced according to the invariant mass method \cite{bibitem37}. The excitation energy of the resonance states of $^{9}$Li decaying into $^{6}$He and $t$ can be obtained from $E_{x} =E_{rel}+E_{thr}$, where $E_{thr}$ is the decay threshold energy. To estimate the resolution of $E_{rel}$ and the geometric detection efficiency, a Monte Carlo simulation was performed considering the energy and spatial resolution of the detectors \cite{bibitem36}. The resolution of $E_{rel}$ is 0.8 MeV at 2.5 MeV and increases to 1.1 MeV at 5.0 MeV. The detection efficiency is simulated over a broad range of relative energies and a typical value at 3.3 MeV is 37$\%$.

\begin{figure}[!htb]
\subfigure{
\label{fig1}
\includegraphics[width=\hsize]{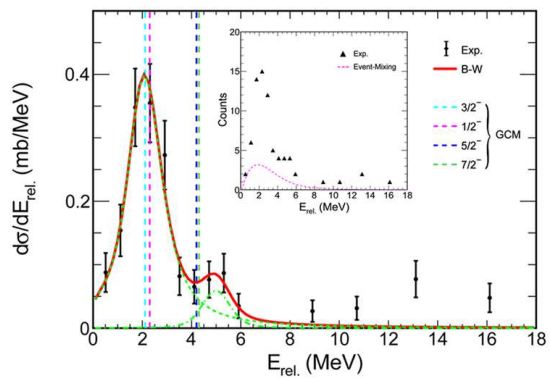}
}
\caption{(color online) The differential breakup cross section $E_{rel}$ spectrum with double Breit-Wigner shaped resonance functions fitting (red line), reconstructed from the $^{6}$He+$t$ coincident fragments, with the nonresonance contribution and acceptance taken into account. The vertical color-lines are the guides for the GCM calculation from Refs. \cite{bibitem25,bibitem26}. Within the inserted figure, the original experimental $E_{rel}$ distribution and the nonresonance contribution estimated by event-mixing method are shown.}
\label{fig1}
\end{figure}
%

The relative energy spectrum of $^{6}$He+$t$ decayed from the excited $^{9}$Li is shown in Fig. 1 and two peaks at 2.2 MeV and 5.0 MeV are well observed. The corresponding excitation energy of $^{9}$Li are 9.8 MeV and 12.6 MeV respectively. They are well agree with the GCM calculation \cite{bibitem25}, the 1/2$^{-}$ and 3/2$^{-}$ states with orbital angular momentum L=1 might be the candidates of the first peak and the 5/2$^{-}$ and 7/2$^{-}$ states with L=3 might be the candidates of the second peak. Taking into account of the intrinsic spin of $^{6}$He (0$^{+}$) and $t$ (1/2$^{+}$), the total angular momentum of $^{6}$He+$t$ resonance states should be contributed by the coupling of the angular momenta of the clusters and the orbital angular momentum of the intercluster motion.

In Fig. 1, the direct (nonresonant) breakup or phase-space distribution is unavoidable for $E_{x}$ just above the decay threshold and can be simulated by implementing the "event-mixing" technique \cite{bibitem38,bibitem39,bibitem40}, and thus should be considered when analyzing resonant states. Together with an "event-mixing" term and the detection acceptance, the double Breit-Wigner (BW) shaped resonance functions convoluted with the detection resolution are used to fit the shape of the $E_{rel}$ spectrum to evaluate their formation and decay natures \cite{bibitem41,bibitem42,bibitem43}. The extracted width of BW function for the first peak is $\Gamma=1.1(2)$ MeV (the error is statistical only) through a least square fitting.

The angular momentums of the resonance peaks at 9.8 MeV and 12.6 MeV can be analyzed by applying the angular correlation method for the breakup reaction described as a(A, B* $\rightarrow$ c+C)b \cite{bibitem44}. Due to the nonzero spins of projectile and the outgoing particle $t$, a more general formalism of angular correlation analysis have to be employed. The calculated angular correlations together with experimental angular correlated spectrum reconstructed from the $^{6}$He+$t$ decay channel respect to x (x=$sin(\Psi)$, with estimated uncertainty of 0.1), in the reaction plane, are shown in Fig. 2, where the $\Psi$ is the c.m. angle relative to the beam direction. The calculated angular correlations are normalized and multiplied by detection acceptance. Because of the detector setup used in the present experiment, the formation transition amplitudes at zero degrees can be used for the present analysis of angular correlation, which are obtained based on CDCC calculation (details are given below). In Fig. 2 (a), the data are gated on a $E_{x}$ range of 9.3 MeV to 11.7 MeV in order to reduce contamination from the event-mixing component and higher energy peaks. Based on the comparison in fig. 2 (a), J$^{\pi}$=3/2$^{-}$ component is mostly consistent with the experimental data for the resonance energy at 9.8 MeV. Nevertheless, if taking the evidence  only  based on present analysis of angular correlation, as shown in Fig. 2 (a), 1/2$^{-}$, 1/2$^{+}$, and 3/2$^{+}$ might also be the candidates for this resonance peak and more proofs needed. Together with the GCM and CDCC calculations, 3/2$^{-}$ state can be considered to be corresponding to the peak at 9.8 MeV. The same method is used to analyze the data within the gated range of 12.0 MeV to 14.6 MeV. As shown in Fig. 2 (b), the angular correlation distribution of 7/2$^{-}$ component can be considered to be mostly in accordance with the experimental data for the resonance energy at 12.6 MeV. Considering the results of GCM calculation based on the deformed $^{6}$He cluster configuration with a specific orientation, the excitation energy of 11- to 14-MeV region might correspond to a mixture of the L=1 component of deformed $^{6}$He (2$^{+}$) with a specific orientation and the L=3 component with $^{6}$He being not deformed  \cite{bibitem25,bibitem26}.

\begin{figure}[!htb]
\subfigure{
\label{fig2}
\includegraphics[width=\hsize]{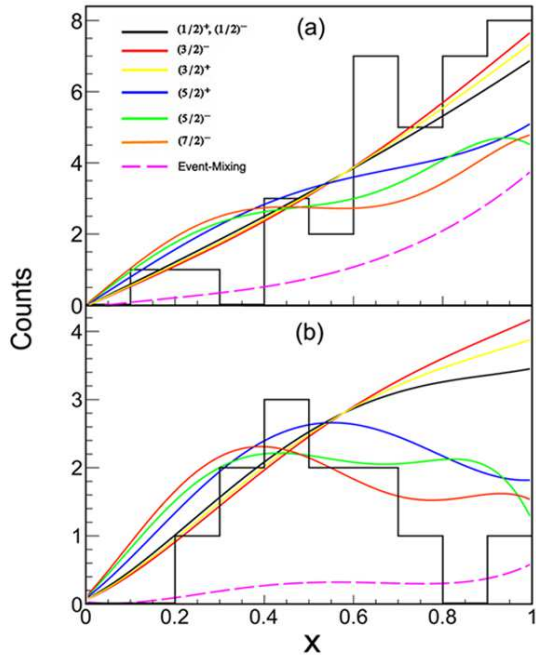}
}
\caption{(color online)  Comparison of calculated angular correlation distributions with experimental data (histogram) reconstructed from the $^{6}$He+$t$ decay channel. (a) for the first peak at 9.8 MeV and (b) for the second peak at 12.6 MeV.}
\label{fig2}
\end{figure}


In Ref. \cite{bibitem25}, $^{6}$He+$t$ cluster states were suggested at the energy region a few MeV higher than its threshold energy. The 1/2$_{2}^{-}$, 3/2$_{3}^{-}$, 5/2$_{2}^{-}$, and 7/2$_{2}^{-}$ states were considered to be members of a K$^{\pi}$=1/2$^{-}$ band. The 1/2$_{2}^{-}$ and 3/2$_{3}^{-}$ states corresponding to L=1 agree with the first cluster resonance peak of the present work. As suggested in the Ref. [26], $^{6}$He(0$^{+}$)+$t$ and $^{6}$He(2$^{+}$)+$t$ decays of $^{9}$Li (3/2$^{-}$) were considered as P-wave decays in three channels, [0$\otimes$3/2]3/2, [2$\otimes$3/2]3/2, and [2$\otimes$1/2]3/2. The $^{9}$Li(3/2$_{3}^{-}$) state is dominated by the $^{6}$He(0$^{+}$)+$t$ component. The $^{9}$Li(3/2$_{4}^{-}$) and $^{9}$Li(3/2$_{5}^{-}$) states can be regarded as $^{6}$He(2$^{+}$)+$t$ cluster resonances because they have dominant $^{6}$He(2$^{+}$)+$t$ components. The $^{6}$He(0$^{+}$)+$t$ component is concentrated at excitation energy of 9-10 MeV. Whereas the $^{6}$He(2$^{+}$)+$t$ components are significant in the excitation energy of 11-14 MeV region.

\section{COUPLED CHANNELS CALCULATIONS AND DISCUSIONS}

The aforementioned analysis demonstrate that, $^{6}$He+$t$ cluster structure with the excitation energies just above the threshold energy might appear in $^{9}$Li. For further investigation, a CDCC calculation with $^{9}$Li modeled as $^{6}$He+$t$ cluster configuration was performed by the code FRESCO \cite{bibitem45}. Nuclear projectiles having a prominent two body structure may undergo breakup by interaction with the nuclear and Coulomb field of a target. The CDCC calculation provides a full quantum mechanical framework to explain the breakup process \cite{bibitem46}. The S\={a}o Paulo (SP) potential connected with the double-folding procedure \cite{bibitem47,bibitem48} is applied to the nuclear interaction between core ($^{6}$He) and target ($^{208}$Pb) and between valence ($t$) and target. The $^{6}$He+$t$ real binding potential consists of a central term and spin orbit term of Woods-Saxon (WS) optical model potential. The central part ($V_{0}$=140 MeV, $r_{v}$=0.80 fm, $a_{v}$=0.66 fm) is obtained from analogy of the parameters applied to $^{6}$Li+$t$ potential of bound state as mentioned in \cite{bibitem49}. The $a_{v}$ is extracted from Ref. \cite{bibitem50}, and references therein, and $r_{v}$ is adjusted to fit the root mean square matter radius of $^{9}$Li in ground state \cite{bibitem51}. The global triton optical model potential discussed in ref. \cite{bibitem52} is applied for spin-orbit term, which shows a very little effect.

In the present calculation, the relative orbital angular momentum for $^{6}$He+$t$ configuration is considered up to L=3 and the $^{6}$He+$t$ continuums are discretized into 20 equally spaced bins by considering $\varepsilon_{max}=20$ MeV. Inclusion of L=4 and L=5 in the calculation has shown less influence on the obtained result. The differential breakup cross section as a function of relative energy for various J$^{\pi}$ values are shown in Fig. 3. The distribution for 3/2$^{-}$ state,in comparison to others, indicates a pronounced resonance peak around the relative energy of 2.0 MeV. This result supports the existence of resonance state at 9.8 MeV, which is also in consonance with the GCM calculation. The cross section for 7/2$^{-}$ state also gives a relative strong cross section with a very wide distribution. Both the angular correlation analysis and GCM calculation indicate the existence of 7/2$^{-}$ state in the excitation energy of 11- to 14-MeV region. It can be seen that the present CDCC calculation do not show enhanced cross section for positive-parity states. In addition, as shown in Fig. 4, within the corresponding $E_{x}$ range of 9.2$-$11.3 MeV, the angular distribution of the differential cross section for 3/2$^{-}$ state with summing over energy bins within the gate, as a function of polar angle (in center of mass frame), well reproduce the distribution of experimental points $(\frac{d\sigma(\theta)}{d\Omega(\theta)})_{exp}$, which is reconstructed from the coincident $^{6}$He+$t$ events. Therefore, the 3/2$^{-}$ state with L=1 as a member of  K$^{\pi}$=1/2$^{-}$ band can be determined to has a structure of $^{6}$He+t resonance with excitation energy at 9.8 MeV.

 A DWBA calculation by assuming a pure 3/2$^{-}$ final state excited from the ground state based on the code FRESCO was performed to describe this inelastic excitation of $^{9}$Li on $^{208}$Pb target, which also well reproduces the experimental differential cross section angular distribution ($\ell$=0 curve shown in Fig .4). This DWBA procedure is based on the multipole-decomposition (MD) analysis of the inelastic differential cross sections in the form of $(\frac{d\sigma(\theta)}{d\Omega(\theta)})_{exp}=\sum_{\ell}a_{\ell}\cdot(\frac{d\sigma(\theta)}{d\Omega(\theta)})_{\ell,DWBA}$ \cite{bibitem53,bibitem54}, where $(\frac{d\sigma(\theta)}{d\Omega(\theta)})_{\ell,DWBA}$ are the DWBA cross sections exhausting the full energy-weighted sum rule (EWSR) for the transferred angular momentum $\ell$. A set of normalization factors $a_{\ell}$, corresponding to the fraction of the EWSR for a monopole ($\ell$=0) transition or a multipole ($\ell$$\geq$1) transition, are applied to fit the data.

\begin{figure}[!htb]
\subfigure{
\label{fig3}
\includegraphics[width=\hsize]{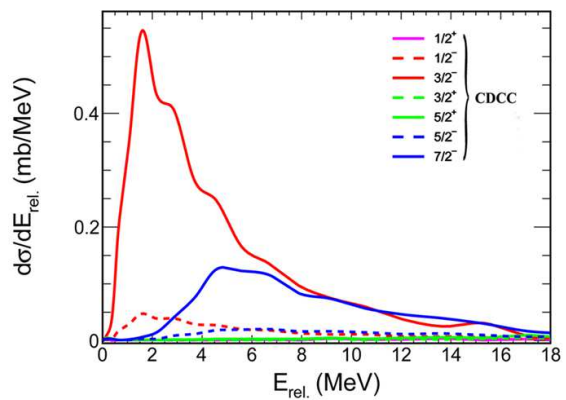}
}
\caption{(color online) The differential breakup cross-section relative energy spectrum for various partial waves (L$\leq$3) from CDCC calculation.}
\label{fig3}
\end{figure}

\begin{figure}[!htb]
\subfigure{
\label{fig4}
\includegraphics[width=\hsize]{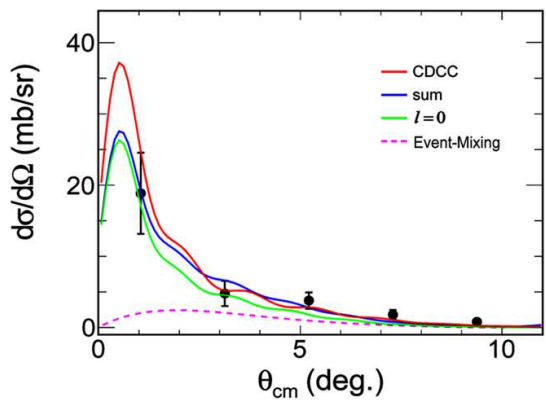}
}
\caption{(color online) Differential cross section angular distribution of 3/2$^{-}$ state (9.8 MeV). The solid red curve is the CDCC calculation. The solid green curve shows the $\ell$=0 contribution. The solid blue curve is the sum of the $\ell$=0 contribution and the event-mixing term.}
\label{fig4}
\end{figure}


In order to extract the isoscalar monopole transition matrix element ($M$(IS0)), the transferred angular momentum $\ell$= 0 and 2 components and the event-mixing term should be taken into account in MD analysis. Nevertheless, due to the CDCC result of that the $E_{x}$ spectrum is dominated by 3/2$^{-}$ state with negligible contaminations from states with other spins and the differential cross section angular distribution displays a dominated monopole excitation, thus only the $\ell$= 0 component is considered in the fitting procedure together with the event-mixing component and the detection acceptance. The theoretical curve is convoluted with the angular resolution (FWHM) of $\sim$0.6 degree obtained from Monte Carlo simulation contributed from uncertainties in determining the reaction position and the position and energy of the fragments. The transition potential for each multipole transition is based on the prescription of the deformed potential model as given in the Refs. \cite{bibitem31,bibitem55,bibitem56}. An equivalent local WS potential to the SP potential connected with the double-folding procedure \cite{bibitem47,bibitem48} is applied to extract the transition potentials.

As shown in the Fig. 4, the $\ell$=0 distribution obtained by the fitting procedure gives a good reproduction of the experimental points, which is also consistent with the CDCC result. The normalization factor $a_{0}$=0.0057(13) is determined by minimizing $\chi^{2}$. This factor is multiplied by 1.5 to account for the fraction of events belong to the 3/2$^{-}$ peak but outside the applied energy gate (9.3$-$11.7 MeV) and thus it is corrected to be 0.0086(20). The deduced $M$(IS0) is estimated to be 4 fm$^{2}$ based on the corresponding EWSR of 4444.12 MeV$\cdot$fm$^{4}$. The uncertainties are mainly from the modeling of the shapes of the 9.8 MeV peak, the DWBA calculation, the event-mixing term and the drop of the $\ell$=2 contribution in the MD analysis procedure.

In the present work, the extracted $M$(IS0) for the 9.8 MeV state of unstable nucleus $^{9}$Li is comparable to those for the typical cluster states in $^{12}$C, $^{16}$O, and unstable nucleus $^{12}$Be and a rough estimated single particle strength 3.6 fm$^{2}$ of $^{9}$Li \cite{bibitem57,bibitem58,bibitem59,bibitem13}. From the GCM calculation for $^{9}$Li with $^{6}$He+$t$ configuration in \cite{bibitem26}, a value of $\sim$6 fm$^{2}$ for $M$(IS0) at excitation energy of 9.8 MeV was obtained by distance parameter D$\leq$15 fm calculation. This is in close agreement with our present observation. As a consequence of the GCM calculation, the strengths for $^{6}$He+$t$ cluster resonances in $^{9}$Li are weakly enhanced because of the large fragmentation of strengths in the corresponding energy region than those obtained for $^{6}$He+$\alpha$ cluster resonances of $^{10}$Be(0$_{3,4}^{+}$) above the $\alpha$-decay threshold. That might be a reason why the extracted $M$(IS0) for unstable nucleus $^{9}$Li is not much enhanced when comparing to that (7 fm$^{2}$) for unstable nucleus $^{12}$Be. This implies that the cluster structure of $^{6}$He(0$^{+}$)+$t$ might be a developed state just above the threshold energy without cluster formation in ground state.

\section{Conclusions}
In summary, $^{6}$He+$t$ cluster resonance states in the breakup reaction of $^{9}$Li on a $^{208}$Pb target have been observed at 32.7 MeV/nucleon. The experiment was carried out by using the $\Delta$E-E telescope detector system around zero degrees at HIRFL-RIBLL. Based on the angular correlation analysis, GCM calculation and CDCC calculation, the spin-parity of $^{6}$He+$t$ resonance states are identified to be 3/2$^{-}$ and 7/2$^{-}$ for the peak at 9.8 MeV and at 12.6 MeV respectively. The two states can be considered as members of K$^{\pi}$=1/2$^{-}$ band. In addition, a comparable monopole transition matrix element to those in $^{12}$C, $^{16}$O, and $^{12}$Be is deduced for excitation state at 9.8 MeV, supporting the picture of strong clustering in excitation states of unstable nucleus $^{9}$Li. States in the excitation energy of 11- to 14-MeV region need to be further investigation. Due to the two clusters $^{6}$He and $t$ are neutron-rich, the cluster formation in $^{9}$Li might be strongly promoted by the redundant neutrons. Further experimental study of cluster states in $^{9}$Li is still desired for understanding the underlying physics of cluster formation in neutron-rich unstable nuclei.
\section{Acknowledgments}
This work was financially supported by the National Natural Science Foundation of China with Grant No. U1432247 and 11575256 and the National Basic Research Program of China (973 Program) with Grant No. 2014CB845405 and 2013CB83440x. We gratefully acknowledge Prof. Y. Kanada-En¡¯yo, Prof. Y. L. Ye, Prof. C. J. Lin and Prof. D. Y. Pang for the discussion. J. Lubian thanks CNPq and FAPERJ for the partial financial support trough the project INCT-FNA Proc. No. 464898/2014-5.

\end{spacing}
\end{document}